\documentclass[twocolumn]{aastex631}
\graphicspath{{./fig/}}

\begin{document}

\title{
Super-Earth formation with slow migration from a ring in an evolving peaked disk compatible with terrestrial planet formation
}

\author[0000-0002-8300-7990]{Masahiro Ogihara}
\affiliation{Tsung-Dao Lee Institute, Shanghai Jiao Tong University, 1 Lisuo Road, Shanghai 201210, China}
\affiliation{School of Physics and Astronomy, Shanghai Jiao Tong University, 800 Dongchuan Road, Shanghai 200240, China}

\author[0000-0001-8476-7687]{Alessandro Morbidelli}
\affiliation{Collège de France, 11 Place Berthelot, 75231 Paris, France}
\affiliation{Laboratoire Lagrange, Observatoire de la Côte d’Azur, Université Côte d’Azur, CNRS, 06304 Nice, France}

\author[0000-0002-1932-3358]{Masanobu Kunitomo}
\affiliation{Department of Physics, Kurume University, 67 Asahimachi, Kurume, Fukuoka 830-0011, Japan}
\affiliation{Laboratoire Lagrange, Observatoire de la Côte d’Azur, Université Côte d’Azur, CNRS, 06304 Nice, France}



\begin{abstract}

For the origin of the radially concentrated solar system's terrestrial planets, planet formation from a ring of solids at about 1 au from the Sun with convergent/suppressed type I migration is preferred. On the other hand, many super-Earths and sub-Neptunes are found in the close-in region with orbital periods of 10--100 days, so that planet formation from rings in the 1-au region would require some degree of inward migration.
One way to realize these different formation scenarios is to use different gas disk models. In this study we investigate whether different scenarios can be realized within a single framework.
We consider a disk model that evolves via disk winds and develops a density peak, and study planet formation and orbital evolution using \textit{N}-body simulations.
Planets with masses less than an Earth mass formed from a low-mass ring resembling the solar system do not migrate inward even in the evolving disk and remain near 1-au orbits, maintaining a high radial mass concentration. On the other hand, planets with masses greater than an Earth mass formed from a massive ring slowly migrate inward above the outward migration region.
As a result, the innermost planet can move to an orbit of about 10 days.
The simulation results also reproduce the characteristics (e.g., mass distribution, eccentricity, orbital separation) of the solar system and super-Earth/sub-Neptune systems.
Our model predicts that Earths and sub-Earths formed by migration from rings at near the 1-au region are less abundant in the close-in region.
\end{abstract}


\section{Introduction}\label{sec:intro}

One of the major characteristics of the terrestrial planets in the solar system is that their masses are concentrated at a radial distance of $r\simeq$ 1\,au from the Sun \citep{2001Icar..152..205C}. There are no planets inside the orbit of Mercury, and the masses of the two planets in the middle (i.e., Earth and Venus) are greater than those of the other planets. To explain the origin of this radially concentrated mass distribution, it is reasonable to assume that the solar system's terrestrial planets formed from planetary embryos with a narrow ring configuration \citep{2009ApJ...703.1131H}.

Terrestrial planet formation from a protoplanet ring is supported by studies of planet formation in earlier phases. Planetesimals can form in rings near $r=1\,{\rm au}$ by the pebble pileup \citep[e.g.,][]{2016A&A...594A.105D}, dead zone inner boundary \citep[e.g.,][]{2019ApJ...871...10U,2021ApJ...921L...5U}, silicate sublimation line \citep[e.g.,][]{2022NatAs...6...72M,2022NatAs...6..357I}, and convergent radial drift in a disk with a density peak \citep{2016A&A...596A..74S,2021ApJ...909...75T}. Planetesimals can also move due to radial drift towards $r \simeq $ 1\,au and form a ring distribution, depending on the density structure of protoplanetary disks \citep{2018A&A...612L...5O}.

Previous study that considered the formation of terrestrial planets from an embryo ring does not consider the effect of orbital changes due to the gas disk \citep{2009ApJ...703.1131H}. However, since the growth time to the mass that initiates the type I migration ($M \sim 0.1\,M_\oplus$) is considered to be shorter than the disk lifetime, the effect of orbital change due to gas should be considered. It has been shown that protoplanets rapidly migrate inward when a simple power-law density distribution is considered \citep[e.g.,][]{2008ApJ...673..487I,2015A&A...578A..36O}. As a result, it is difficult to explain the radial mass concentration of the solar system's terrestrial planets. A possible solution to this problem is to consider a gas disk model with a density peak near $r=$ 1\,au \citep[e.g.,][]{2021NatAs...5..898B,2023Icar..39615497W}. Such a disk model would result in a concentration of protoplanet orbit around $r=$ 1\,au as a consequence of convergent type I migration. This is because in regions where the local gas surface density gradient is positive, the barotropic part of the corotation torque becomes positive and overwhelms the other torques, thus realizing outward migration \citep{2006ApJ...642..478M,2010MNRAS.401.1950P}.

In fact, recent studies of protoplanetary disks have pointed out that the actual disk structure may not be expressed by the simple power-law distribution previously thought. Gas disks with surface density peaks near $r=$ 1\,au can be realized when the evolution by magnetically driven disk winds is considered. \citet{2016A&A...596A..74S} modeled the mass loss and angular momentum transport due to disk winds from MHD simulations and calculated the long-term evolution of the 1D density profile. The results show that, depending on the efficiency and radial profile of wind-driven accretion, a disk density structure with a surface density peak near $r=$ 1\,au can be obtained \citep[see also][]{2022MNRAS.512.2290T}\footnote{As mentioned above, this could cause a convergent radial drift of dust, creating a planetesimal ring near $r=1{\rm \,au}$.}.
As another model, a disk with a density peak near $r=1\,{\rm au}$ can also be realized when the accretion rate changes gradually at the dead zone inner boundary \citep{2022MNRAS.509.5974J}.

Meanwhile, the orbital distribution of the low-mass planets in exoplanet systems, super-Earths and sub-Neptunes (hereafter referred to as SENs), is different from that of the solar system. Although there are some observational limitations, typical masses of SENs are $M \simeq 1-10\,M_\oplus$, with semimajor axis ranging from $a =$0.1--1\,au.
More detailed information on mass distribution, occurrence rate, orbital separation, etc. is also available \citep{2023ASPC..534..863W,2023ASPC..534..839L}.
Thanks to these information, our understanding of the origin of SENs has improved dramatically in the past five years or so \citep[][]{2015A&A...578A..36O,2017MNRAS.470.1750I,2021A&A...650A.152I,2019A&A...627A..83L}. The fact that orbits of most SENs are not in mean-motion resonances and that they do not have ice-rich compositions suggests that these planets did not experience a large-scale orbital migration over several au. Nevertheless, some orbital migration of about $\sim 1$\,au or less in a long timescale of about 1\,Myr is consistent with observations \citep{2018A&A...615A..63O,2020ApJ...899...91O,2019A&A...627A..83L,2023ASPC..534..863W}.
It has been proposed that super-Earth cores form from narrow rings \citep{2023NatAs...7..330B} and wide rings \citep{2019A&A...627A..83L} near $a = 1\,{\rm au}$, and then migrate to $a\simeq$ 0.1\,au in disks with a power-law surface density distribution, radically different from those invoked for terrestrial planet formation in the solar system.

Thus, we seem to face a contradiction:
Convergent or suppressed migration is favorable for solar system formation, and inward migration seems to be favorable for SEN formation. Conditions like those in \citet{2023Icar..39615497W}, where convergent migration works, do not explain the close-in orbits of SENs, and conditions like those in \citet{2023NatAs...7..330B}, where inward migration is prominent, do not explain the radial mass concentration of the solar system. To realize these separate scenarios, separate disks, either solid or gaseous, must be considered.
Therefore, in this study, we investigate whether the same peaked gas disk model can explain the origin of both the solar system and SEN systems. Specifically, we focus on the possibility of inward migration even in the case of peaked disks, depending on the saturation of the corotation torque.

This paper is organized as follows. In Section\,\ref{sec:model}, we describe the model and give the setting of \textit{N}-body simulation.
In Section\,\ref{sec:results}, we present our typical results.
In Section\,\ref{sec:characteristics}, we show some important properties of the simulated systems and compare them with those for the solar system and observed SEN systems.
In Section\,\ref{sec:discussion}, we present a discussion on the prediction of the occurrence of Earths and sub-Earths, and on the correlation between planets and metallicity.
In Section\,\ref{sec:conclusions}, we summarize our conclusions.

\section{Model}\label{sec:model}

\subsection{Disk model and migration}\label{model:disk}

\begin{figure*}[ht!]
\plotone{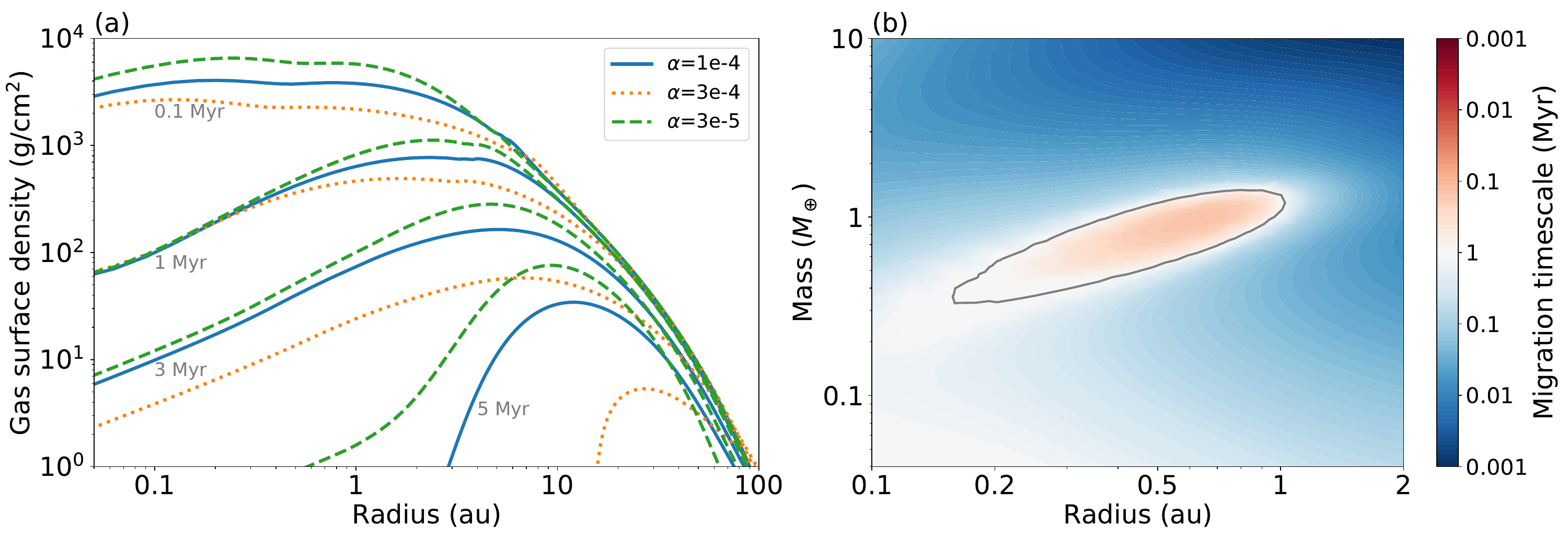}
\caption{(a) Temporal evolution of the gas surface density. Different lines indicate disks with different $\overline{\alpha_{r\phi}}$ parameter.
(b) Migration map for the disk with $\overline{\alpha_{r\phi}} = 10^{-4}$ at $t = 1\,{\rm Myr}$. The colors indicate the direction of migration, with red indicating outward migration and blue inward migration. The solid line shows the zero migration curve, which separates the regions of inward and outward migration. Note that the map depends on the eccentricity and inclination, which are set to 0.005.
\label{fig:disk}}
\end{figure*}

The evolution of the gas surface density is obtained by solving the 1D diffusion equation, where the angular momentum transport due to viscosity and magnetically driven disk winds (wind-driven accretion) and the mass loss due to magnetically driven disk winds and photoevaporation are included. For details, readers are referred to \citet{2016A&A...596A..74S} and \citet{2020MNRAS.492.3849K}.
\begin{eqnarray}\label{eq:sgmevl}
\frac{\partial \Sigma_{\rm g}}{\partial t} &=& \frac{1}{r}\frac{\partial}{\partial r}
\left\{\frac{2}{r\Omega}\left[\frac{\partial}{\partial r}(r^2 \Sigma_{\rm g}
  \overline{\alpha_{r\phi}}c_{\rm s}^2) + r^2 \overline{\alpha_{\phi z}}
  \frac{\Sigma_{\rm g}H\Omega^2}{2\sqrt{\pi}} \right]\right\} \nonumber\\
&& - \dot{\Sigma}_{\rm MDW} - \dot{\Sigma}_{\rm PEW}
\end{eqnarray}
where $\Sigma_{\rm g}, \Omega, c_{\rm s},$ and $H$ are the gas surface density, the angular velocity, the sound velocity, and the scale height, respectively.
The first term on the right hand side represents the viscous accretion, which is determined by the parameter $\overline{\alpha_{r\phi}}$. For this parameter, we consider $\overline{\alpha_{r\phi}} = 10^{-4}$ as fiducial, because recent theoretical and observational studies favor weak turbulence \citep[e.g.,][]{2017ApJ...843..150F,2018ApJ...869L..46D}. We also consider cases where the parameter is increased/decreased by a factor of three.
The second term on the right hand side is the wind-driven accretion, which is controlled by the parameter $\overline{\alpha_{\phi z}}$. For this parameter, we use the relation $\overline{\alpha_{\phi z}}=\min[10^{-5} (\Sigma_{\rm g}/\Sigma_{\rm g,ini})^{-0.66},1]$, which is considered to increase (the plasma beta decreases) with disk dissipation (see Eq.\,(30) of \citealt{2016A&A...596A..74S}).
The disk temperature is determined by stellar irradiation and viscous heating as in \citet{2020MNRAS.492.3849K}.

The mass loss due to the magnetically driven disk winds is given by
\begin{equation}
    \dot{\Sigma}_{\rm MDW} = C_{\rm w} \frac{\Sigma_{\rm g}\Omega}{2\sqrt{\pi}}.
\end{equation}
The parameter $C_{\rm w}$, which represents the strength of disk winds, is given as $C_{\rm w} = \min(C_{\rm w,0}, C_{\rm w,e})$. We use $C_{\rm w,0}=10^{-5}$ suggested by MHD simulations \citep[e.g.,][]{2010ApJ...718.1289S,2014ApJ...784..121S} for the weak turbulence case. The constraint of $C_{\rm w,e}$ comes from energetics, and in this study we consider the ``weak DW case'' in Eq.\,(20) of \citet{2016A&A...596A..74S}, which is a conservative limit. Here, the maximum energy of wind particles is considered to be 10\% of the available energy, which is the sum of the gravitational energy released by accretion and the energy released by viscous heating.
For the mass loss due to photoevaporation, $\dot{\Sigma}_{\rm PEW}$, we use models based on the results of hydrodynamical simulations \citep{2007MNRAS.375..500A,2012MNRAS.422.1880O}. We consider $10^{30}$ erg\,s$^{-1}$ as the X-ray luminosity and $10^{41}\,{\rm s^{-1}}$ as the EUV photon flux. See Section\,2.4 of \citet{2020MNRAS.492.3849K} for specific formulas.

Figure~\ref{fig:disk}(a) shows the time evolution of the disk gas surface density. The initial condition for the gas surface density is the power-law distribution proportional to $r^{-2/3}$ as in the minimum mass solar nebula with exponential cut-off. A relatively massive gas disk with a disk mass of $0.118\,M_\odot$ is considered as the initial condition to connect with the early stage of disk evolution. To investigate the later stages of planet formation in our \textit{N}-body simulations, the times used in Figure~\ref{fig:disk}(a) are delayed by 0.1 Myr from the disk evolution simulations.

As can be seen in the figure, the gas surface density decreases in the inner region of the disk due to wind mass loss and wind-driven accretion. Around $t=0.1 {\rm \,Myr}$ (i.e., 0.2 Myr after the start of the disk evolution simulation), the slope of the surface density is almost flat within $r\simeq$ 1 au, and then the slope is positive. After $t\simeq 5 {\rm \,Myr}$, the inner disk is rapidly cleared by photoevaporation. The distribution at $t=1 {\rm \,Myr}$ is similar to the ``shallow disk'' model of \citet{2023Icar..39615497W}, which did no include any time-evolution of the disk other than a uniform gradual depletion.

Planets and planetary embryos in the disk are subject to orbital migration and eccentricity and inclination damping as a result of gravitational interaction with the disk gas.
The equation of motion of particle $i$ at position $\mathbf{r}_i$ is given by
\begin{eqnarray}
\frac{d^2 \mathbf{r}_i}{dt^2} &=& - \frac{GM_* \mathbf{r}_i}{|\mathbf{r}_i|^3} - \sum_{j \ne i} \frac{GM_j (\mathbf{r}_i - \mathbf{r}_j)}{|\mathbf{r}_i - \mathbf{r}_j|^3} - \sum_j \frac{GM_j \mathbf{r}_j}{|\mathbf{r}_j|^3} \nonumber \\
 &&+ \mathbf{F}_{\text{mig}} + \mathbf{F}_{\text{damp}} \label{eq:eom}
\end{eqnarray}
where $M_*$ is the stellar mass.
For detailed expressions for the specific forces for the migration and damping ($\mathbf{F}_{\text{mig}}$ and $\mathbf{F}_{\text{damp}}$), the reader is referred to \citet{2015A&A...579A..65O} and \citet{2017MNRAS.470.1750I}. The timescale for the type I migration is given by
\begin{eqnarray}
t_a \simeq t_{\rm wave}
\left(\frac{\Gamma}{\Gamma_0}\right)^{-1}
\left(\frac{H}{r}\right)^{-2},\\
t_{\rm wave} = \left(\frac{M}{M_*}\right)^{-1}
\left(\frac{\Sigma_{\rm g} r^2}{M_*}\right)^{-1}
\left(\frac{H}{r}\right)^4
\Omega^{-1},
\end{eqnarray}
where $\Gamma/\Gamma_0$ is the normalized migration torque. The migration torque $\Gamma$ is the sum of the Lindblad torque and the corotation torque, and the latter can be positive when the local gas surface density gradient, $\partial \ln \Sigma_{\rm g} / \partial \ln r$, is positive \citep[e.g.,][]{2009MNRAS.394.2283P}. However, the angular momentum in the horseshoe region near the planet is finite, and the corotation torque is prone to saturation. Diffusion is necessary to avoid saturation. When the diffusion timescale and the horseshoe libration timescale are comparable, the corotation torque can be large \citep{2011MNRAS.410..293P}. The parameter $P_\nu$, which controls the saturation of the barotropic part of the corotation torque, is given by
\begin{eqnarray}
    P_\nu &=& \frac{2}{3} \sqrt{\frac{r^2 \Omega}{2 \pi \nu} x_{\rm s}^3}, \\
    x_{\rm s} &\simeq& \sqrt{\frac{M}{M_*}\frac{r}{H}},
\end{eqnarray}
where $x_{\rm s}$ is the dimensionless half width of the horseshoe region \citep{2006ApJ...642..478M,2009MNRAS.394.2297P}. It can be seen that the type I migration torque depends on the planetary mass and diffusion. In addition, the corotation torque decreases also with increasing eccentricity \citep{2010A&A...523A..30B}.
Figure~\ref{fig:disk}(b) shows the migration timescale of the disk at $t=1\,{\rm Myr}$. Outward migration, which occurs in the red region, is seen at $r\lesssim 1 {\rm \,au}$ thanks to the positive gas surface density gradient, but the outward migration region also depends on the planetary mass. Therefore, the conditions under which convergent migration occurs are limited.

The timescales for the damping of eccentricity and inclination are given by \citep{2008A&A...482..677C}
\begin{eqnarray}
t_e = \frac{t_{\text{wave}}}{0.780} \left[ 1 - 0.14 \left(\frac{e}{H/r}\right)^2 + 0.06 \left(\frac{e}{H/r}\right)^3 \right. \nonumber \\
\left. + 0.18 \left(\frac{e}{H/r}\right) \left(\frac{i}{H/r}\right)^2 \right],\\
t_i = \frac{t_{\text{wave}}}{0.544} \left[ 1 - 0.30 \left(\frac{i}{H/r}\right)^2 + 0.24 \left(\frac{i}{H/r}\right)^3 \right. \nonumber\\
\left. + 0.14 \left(\frac{e}{H/r}\right)^2 \left(\frac{i}{H/r}\right) \right].
\end{eqnarray}

\subsection{Simulation setting}\label{model:setting}

We investigate the formation of low-mass planets from planetary embryos in an evolving disk with a density peak shown in Fig.~\ref{fig:disk}(a) by performing \textit{N}-body simulations. For reference, some simulations are also performed using a disk model with power-law density profiles. In addition to the mutual gravity between the planets, the effect of the gas disk (orbital migration and $e, i$ damping) is included in the simulation as force terms (see Eq.~(\ref{eq:eom})). 
The orbits of the planets are integrated for about 200 Myr (for solar system formation) or 100 Myr (otherwise) using the fourth-order Hermite scheme with the hierarchical timestepping \citep{1991ApJ...369..200M, 1992PASJ...44..141M}. When planets collide with each other, perfect accretion is considered.

The initial embryo distribution based on the settings and results of previous \textit{N}-body simulations from rings \citep[e.g.,][]{2009ApJ...703.1131H, 2023Icar..39615497W, 2023NatAs...7..330B}, and we place the embryo in the $a=$ 1--1.5\,au ring region. The mass of each embryo is set to $0.05 \,M_\oplus$. The solid surface density through the ring is proportional to $r^{-3/2}$. The central star mass is $1\,M_\odot$. The total embryo mass is treated as a parameter and is considered to be between 2--20\,$M_\oplus$. The initial eccentricity and inclination are assumed to be small $\simeq 0.01$.

\section{Typical results}\label{sec:results}

\begin{figure*}[ht!]
\plotone{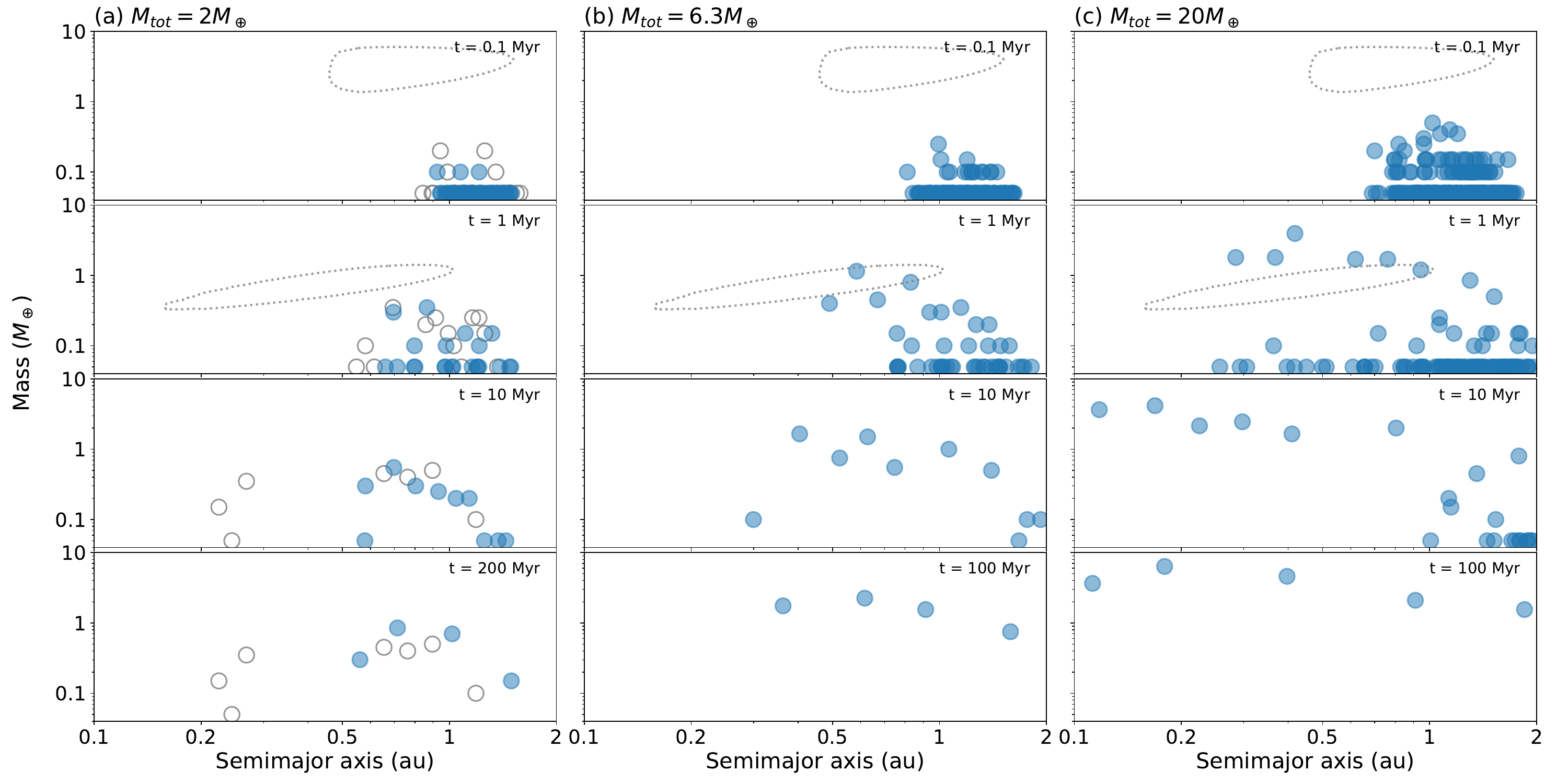}F
\caption{Snapshots of the system from typical simulations. Blue filled circles represent the planets at each time.
(a) The total solid mass is $2\,M_\oplus$ and for solar system formation. The result for the simulation in the power-law disk is shown in gray circles.
(b)(c) The total solid mass is larger and for SEN system formation.
In each panel, the zero-migration curve is also shown in gray dotted lines.
\label{fig:snap}}
\end{figure*}

First, we present typical results with the parameter of $\overline{\alpha_{r,\phi}}=10^{-4}$. Figure~\ref{fig:snap} shows the results for three different initial total solid masses.

Figure~\ref{fig:snap}(a) shows the case where the total mass is small, $M_{\rm tot} = 2\,M_\oplus$. Planetary embryos grow by collisions, and at $t=1\,{\rm Myr}$, the mass of the largest planet is about $0.3\,M_\oplus$. This is roughly in agreement with the results of \citet{2023Icar..39615497W}, which performed higher-resolution simulations. Protoplanets that grow to masses greater than about $0.1 \,M_\oplus$ are affected by type I migration; however, protoplanets do not undergo significant inward migration in the evolving gas disk. It is noteworthy that the radial confinement of embryos can be maintained even in the more realistic time-evolving disk than in the fixed disk with density peak at $r=1\,{\rm au}$ used by \citet{2023Icar..39615497W}. Before $t=1 \,{\rm Myr}$, the density peak at $r=1 {\rm \,au}$ is not pronounced, but at this stage the planetary mass is small and planets do not migrate inward.
The gas in the inner region $r \lesssim 1\,{\rm au}$ disappears by $t=5 {\rm \,Myr}$ (Fig.~\ref{fig:disk}(a)), and before and after that time, protoplanets exhibit giant impacts with each other, eventually forming planets with a maximum mass of about $1\,M_\oplus$. At the end of Fig.~\ref{fig:snap}(a), four planets remain within the orbits of Mercury and Mars. We performed a total of ten simulation runs for each condition, and obtained qualitatively similar results in the other nine runs (see Section~\ref{sec:characteristics}).

As a reference, we have performed simulations in a gas disk with power-law density profiles (two-component power-law disk in which the disk temperature is determined by viscous heating and stellar irradiation in the inner and outer regions, respectively) \citep{2015A&A...582A.112B,2019A&A...632A...7L} under the same conditions. Gray points in Fig.~\ref{fig:snap}(a) shows a typical result. We observe that protoplanets experience inward migration and move within Mercury's present orbit. 

Figure~\ref{fig:snap}(b) and (c) show the results for larger initial total masses, $M_{\rm tot} = 6.3\,M_\oplus$ and $20\,M_\oplus$. The larger the total mass, the faster the growth and migration. At an early stage $(t \lesssim 0.1\,{\rm Myr})$, some planets with mass $> 0.1\,M_\oplus$ appear, which are affected by type I migration. Then, around $t=1\,{\rm Myr}$, some planets grow to masses $> 1\,M_\oplus$, and such planets are prone to rapid migration.
However, in the disk that develops a density peak considered in this study, the direction and speed of orbital migration are affected. In particular, within the region indicated by the zero migration curve in Fig.~\ref{fig:snap}(c), planets can experience the unsaturated positive corotation torque. Thus, planets located inside this region can migrate outward and experience collisions in the convergent region \citep[][]{2020MNRAS.496.3314W,2024A&A...682A..89P}.
However, there is only a limited outward migration region.
As a result, planets with $M=$1--3$\,M_\oplus$ slowly migrate inward above the outward migration region. Note that the migration speed of the planets is slow, because the gas surface density in the inner region is low at $t \sim 1\,{\rm Myr}$, when the planets migrate. The final result in Fig.~\ref{fig:snap}(c) is that more massive planets form in close-in orbits, which is in contrast to the result in Fig.~\ref{fig:snap}(a).

Although not shown here, simulations in the gas disk with a power-law distribution have been performed for the massive initial total mass case $M_{\rm tot} = 20\,M_\oplus$. In this case, planets of about $M\simeq$1-3$\,M_\oplus$ undergo very rapid migration of the order of 0.1\,Myr \citep[][]{2015A&A...578A..36O}. This is considerably faster than the migration speed observed in Fig.~\ref{fig:snap}(c), where the peaked gas profile is considered.

As described above, there is only a limited range of planets that undergo outward/convergent migration in peaked disks, and therefore it is possible for planets to migrate inward in such disks. For a turbulent viscosity of a disk of about  $\overline{\alpha_{r,\phi}}=10^{-4}$, inward migration is dominant for masses about $1\,M_\oplus$ and above. This can naturally explain the difference between the solar system and SEN systems, even when using the same disk model. 
In the next section, we will also look at the simulation results when the  $\overline{\alpha_{r,\phi}}$ is increased/decreased by a factor of three. As we will see, our overall results do not change for this level of difference in the viscosity.

\section{Characteristics of systems}\label{sec:characteristics}
In this section, we look at individual properties of the solar system and SEN systems in more detail, including results of simulations with different disk parameter.

\subsection{Solar system}

\begin{figure*}[ht!]
\plotone{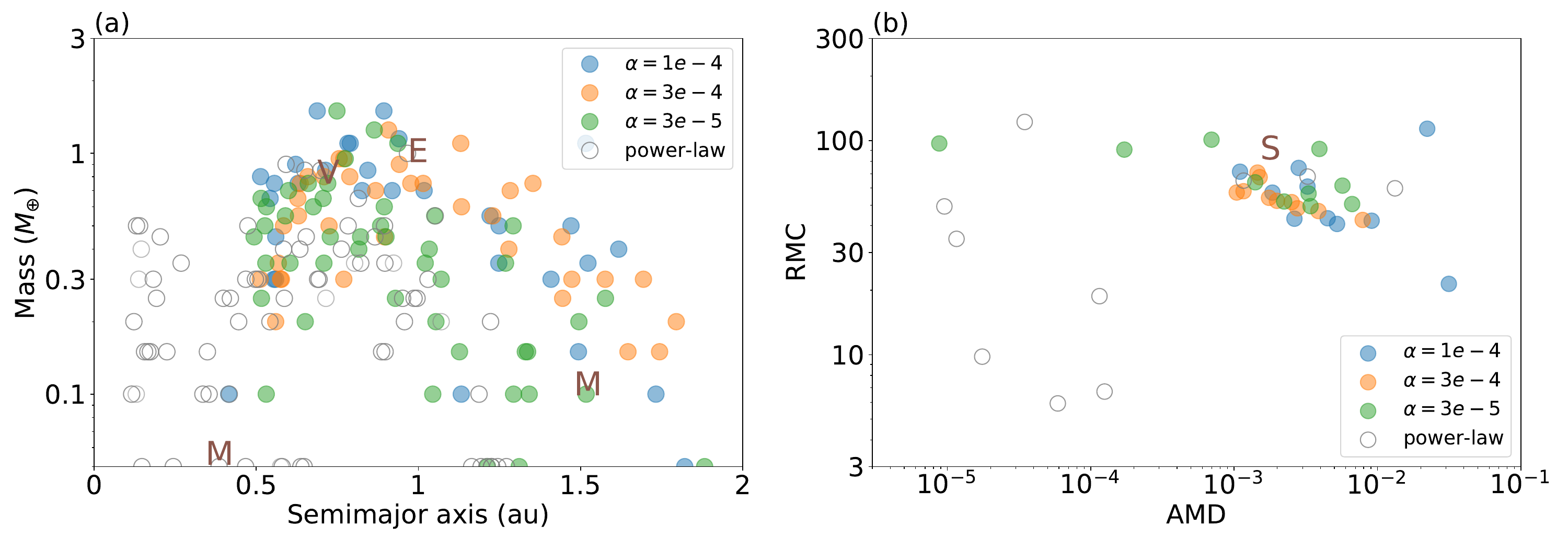}
\caption{Final properties of the simulations with $M_{\rm tot} = 2\,M_\oplus$ for different values of $\overline{\alpha_{r,\phi}}$. Gray symbols indicate results in the power-law disk. Ten simulation runs are performed in each disk.
(a) The final mass distribution. Mercury, Venus, Earth, and Mars are represented by M, V, E, and M, respectively.
(b) The metrics of AMD and RMC in the final state. The terrestrial planets of the current solar system is indicated by S. 
\label{fig:results_ss}}
\end{figure*}

We see whether some important features in the solar system are reproduced in the simulation for the case of small initial total solid mass and $M_{\rm tot} = 2\,M_\oplus$. Figure~\ref{fig:results_ss}(a) is the final mass distribution. For each set, 10 runs of simulations were performed with different initial embryo positions, and all of these results are displayed. The different symbols represent the results performed with disk evolution for slightly different values of the turbulent viscosity, $\overline{\alpha_{r,\phi}}$. Although for $\overline{\alpha_{r,\phi}}=3 \times 10^{-5}$ the mass tends to be slightly smaller in the region of $a > 1\,{\rm au}$ (this will be discussed later), we do not see much difference in results for different values.
The results show that no planets migrate inside the Mercury's orbit, which is remarkable compared to simulations in the power-law disk, where some innermost planets move within $a=0.2 \,{\rm au}$.
The planets near $a=1\,{\rm au}$ have the largest mass, which is approximately $1\,M_\oplus$. Together with the smaller masses of the planets at either end of the orbit, the $a-M$ diagram shows an inverted V shape. From these results, we find that the mass distribution of the final planets is approximately the same as that of the terrestrial planets in the solar system.

The important factor that can explain the mass distribution of the solar system's terrestrial planets is the suppression of orbital migration. It has been shown that planet formation simulations from embryo rings that ignore orbital migration, as in \citet{2009ApJ...703.1131H}, can explain the mass distribution well. On the other hand, when planets experience some degree of orbital migration, as in \citet{2019A&A...627A..83L}, the innermost planets tend to move into inner orbits and become larger than Mercury. In the low total solid mass case of this study, the migration is significantly suppressed in the gas disk with peaked density profile, as seen in Section~\ref{sec:results}.

Figure~\ref{fig:results_ss}(b) shows the values of RMC and AMD, which are metrics that describe the characteristics of planetary systems. The RMC is used as a measure of the degree of radial mass concentration \citep{2001Icar..152..205C}. 
\begin{equation}
{\rm RMC}= \max \left(\frac{\sum_j M_j}{\sum_j M_j (\log_{10} \sqrt{a/a_j})^2} \right).
\end{equation}
The solar system's terrestrial planets have a high mass concentration with the value of 89.9.
In previous simulations, when the simulation is not started from a narrow ring, the final RMC is small, about 50 or less \citep[][]{2009Icar..203..644R,2010Icar..207..517M}. Our results show that the RMC is about 40--100 and reasonably large. This is because planets do not undergo significant orbital migration and the high radial mass concentration given as the initial condition is maintained.
As mentioned in Section~\ref{sec:results}, it is worth noting that the distribution of embryos is kept radially confined even in the evolving disk, rather than always having a fixed density profile.
For the parameter $\overline{\alpha_{r,\phi}}$ dependence, there is not much difference in the results.
Incidentally, in some simulations in the power-law disk, the initial radial mass concentration cannot be maintained due to migration, and the RMC is relatively small $(< 30)$ at the end.

The AMD (Angular Momentum Deficit) is used as a quantitative measure of orbital excitation \citep{1997A&A...317L..75L}.
\begin{equation}
{\rm AMD}= \frac{\sum_j M_j \sqrt{a_j} (1 - \sqrt{1-e_j^2} \cos{i_j})}{\sum_j M_j \sqrt{a_j}},
\end{equation}
where $e_j$ and $i_j$ are the orbital eccentricity and inclination, respectively. The AMD of the solar system's terrestrial planets is 0.0018.
The results of our simulations show that the values for AMD are similar to those of the solar system $(10^{-3}-10^{-2})$, regardless of the values of the disk parameter.
We see AMD values for some systems are slightly higher than that of the current solar system. This is normal, because we only consider embryos in our \textit{N}-body simulations, and there is no damping force such as dynamical friction from planetesimals after disk dissipation.
Note that in some runs for $\overline{\alpha_{r,\phi}}=3 \times 10^{-5}$, the AMD is as small as $\lesssim 10^{-4}$. The dynamical instability after disk dissipation is less pronounced in these runs, leaving smaller planets with small eccentricities that have not experienced many giant impacts\footnote{In one run in particular, ten planets remain small $M<0.4\,M_\oplus$ with low eccentricity at $t=200\,{\rm Myr}$.}.
This also explains the smaller masses in the $a>1\,{\rm au}$ region for $\overline{\alpha_{r,\phi}}=3 \times 10^{-5}$ in Fig.~\ref{fig:results_ss}(a).

On the other hand, the AMD is typically very small for simulations in the power-law disk. In these simulations, planets can be captured in mean-motion resonances during migration, which remains until the final state at $t=200\,{\rm Myr}$. The systems are very stable as a result and do not experience giant impacts after the disk dissipation, which results in many planets with small eccentricities.
It is possible that such systems will experience a giant impact phase later in their long-term evolution $(t>200\,{\rm Myr})$, and thus the AMD would increase to a value similar to that of the solar system. However, in this case, the timing of the giant impact would be later than the timing of the last giant impact experienced by the solar system terrestrial planets \citep[50--150 Myr,][]{2007Natur.450.1206T,2019NatGe..12..696T}, and thus not considered to be consistent with the characteristics of the solar system.

\subsection{SEN system}

\begin{figure*}[ht!]
\plotone{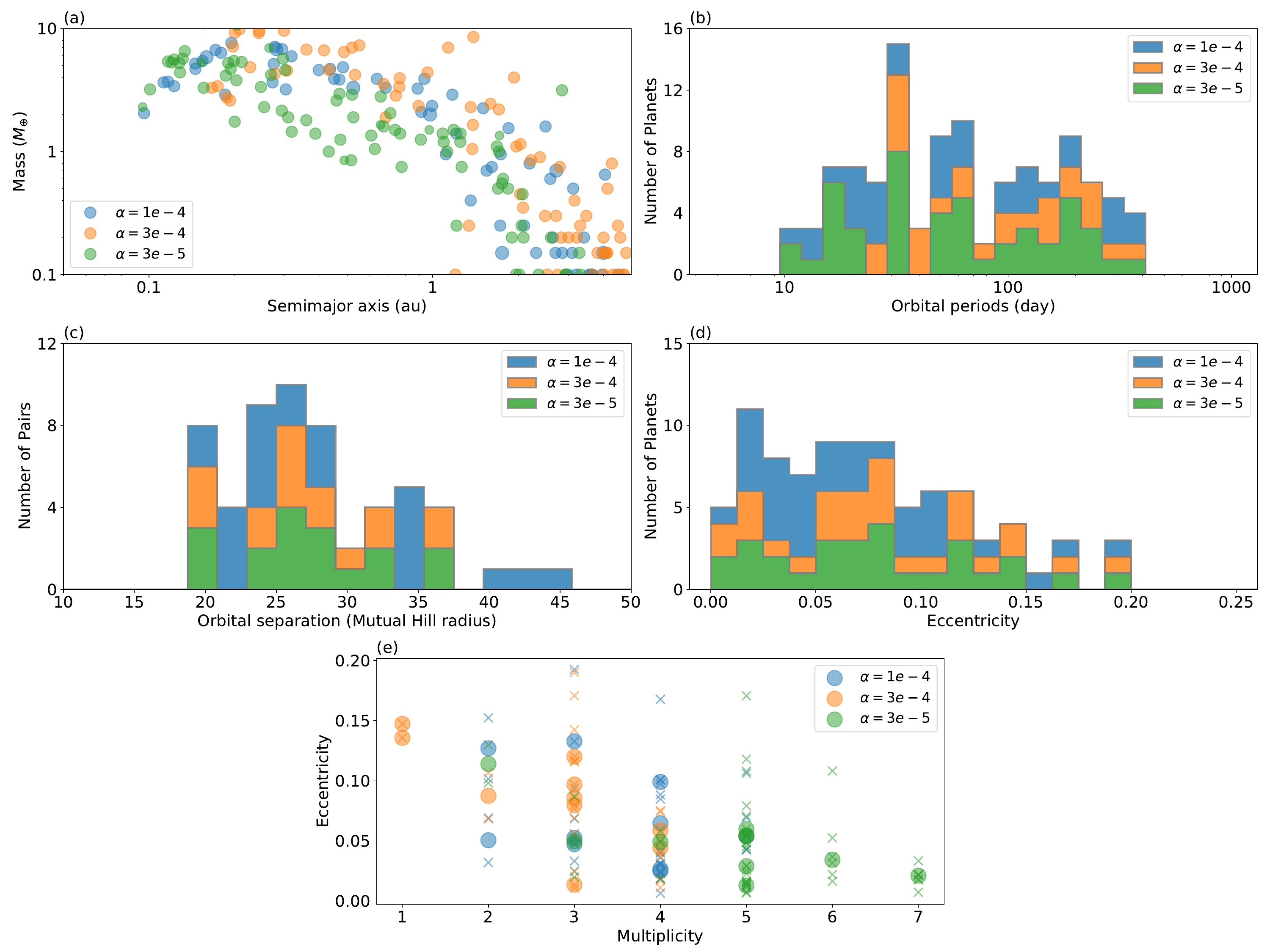}
\caption{Final properties of the simulations with $M_{\rm tot} = 20\,M_\oplus$ for different values of $\overline{\alpha_{r,\phi}}$. Ten simulation runs are performed in each disk. In panels (b)--(e), only planets with $a \leq 1{\rm \,au}$ and $M \geq 1\,M_\oplus$ are considered.
(a) The final mass distribution.
(b) Planet occurrence at each orbital period.
(c) Orbital separation divided by the mutual Hill radius.
(d) Orbital eccentricity.
(e) Relationship between orbital eccentricity and multiplicity. Crosses represent the eccentricity of the planet in each system, and filled circles represent the median in each system.
\label{fig:results_se}}
\end{figure*}

Next, we look at the characteristics of simulations for SEN systems where the initial total mass is set to $M_{\rm tot} = 20\,M_\oplus$. Figure~\ref{fig:results_se} is a summary of the main features.
In panels (b)--(e), we only consider planets with $a \leq 1{\rm \,au}$ and $M \geq 1\,M_\oplus$.

Figure~\ref{fig:results_se}(a) shows the final mass distribution, which exhibits several features.
First, the mass distribution is nearly flat in $a=0.1$ to 1\,au, indicating that the masses are nearly constant. Planetary embryos grow in the ring at $a \simeq 1{\rm \,au}$ and begin to move inward when they reach a critical mass for migration $(M \simeq 1\,M_\oplus)$. Once the planets start migrating, they are out of the ring and do not grow dramatically thereafter (although they do gain a factor of mass in giant impacts). Therefore, there is no steep gradient in the $a=0.1$--1\,au region. The mass is about 1--10$\,M_\oplus$, which is consistent with the SEN mass suggested by the mass-radius relation for Kepler planets \citep[e.g.,][]{2014ApJ...783L...6W,2016ApJ...825...19W}.

Another feature is a trend of decreasing mass beyond the region where the ring of solids initially existed ($a\simeq1{\rm \,au}$). This is a characteristic of planet formation from rings with slow inward migration. 
Note that this decreasing part ($a \gtrsim 1{\rm \,au}$ and $M \lesssim 1\,M_\oplus$) is not observed in current exoplanet observations because it is below the detection limit. This may be found in future observations.
Meanwhile, some observations are finding signs of this trend. \citet{2022AJ....164...72M} has pointed out that for systems with high multiplicity ($\geq 4$) found in Kepler observations, few planets are found in orbits with periods longer than 100 days. This is consistent with our simulation results, i.e. it can be explained by the small mass of the planet, which is not observed at orbital periods beyond 100 days ($a \gtrsim 0.4 {\rm \,au}$). In fact, the planetary masses beyond 100 days are smaller for high-multiplicity systems in our simulations.

As for the difference in the disk parameter, the smaller the $\overline{\alpha_{r\phi}}$ value, the slower the evolution of the disk (Fig.~\ref{fig:disk}(a)), indicating the migration of smaller planets into inner orbits.
We also see that low-mass planets ($M\lesssim 1\,M_\oplus$) do not form in the close-in region ($a\sim 0.1\,{\rm au}$). This is because the type I migration is suppressed for low-mass planets. This will be discussed in Section~\ref{sec:discussion}.

Figure~\ref{fig:results_se}(b) shows the orbital period distribution of the formed planets. Observations show that the occurrence rate of SENs is almost flat in logarithmic bins at about $P=$10--300 days \citep[][]{2018AJ....155...89P,2023AJ....166..122D}. Previous simulations of rapid migration in gas disks with a power-law density distribution \citep[][]{2015A&A...578A..36O,2017MNRAS.470.1750I,2017A&A...607A..67M}, show that SENs pile up near the inner edge of the disk, which does not explain the observed flat occurrence. On the other hand, our occurrence rates are almost flat and consistent with the observation.
The position of the innermost planet is about $P=10$ days, which is also roughly consistent with the observation. The parked position is not determined by the inner edge of the disk, but by the final location of the migration in our simulation. This is because the surface density slope of the gas disk is positive in the close-in region (Fig.~\ref{fig:disk}(a)), and as planets migrate inward, the gas surface density becomes smaller and the migration slows down. This position depends on the initial position of the ring and the disk parameters; therefore it seems that quantitative numbers are not very meaningful. Nevertheless, the fact that there is a cutoff at some point is a general result, and it is consistent with observations.

Figure~\ref{fig:results_se}(c) shows the orbital separation divided by the mutual Hill radius, $R_{\rm H}$. Observations indicate that the orbital separation is about 10--50 $R_{\rm H}$, with 20--30 $R_{\rm H}$ being the majority \citep{2014Natur.513..336L,2023ASPC..534..863W}. Our simulations show that some of the planets are trapped in mean-motion resonances (mostly close resonances such as 4:3 and 5:4) and have rather small orbital separations before the disk dissipates. However, most of the planets undergo giant impacts during and after the disk dissipation and form with large orbital separations of about 20--30 mutual Hill radii. This is in good agreement with observations.
Note that most mean-motion resonances are broken by giant impacts, and many planets eventually have no resonant configurations. Planetary pairs that are in resonant relationships have more separated commensurability (e.g., 2:1) than those in the formation stage in the disk.

Figure~\ref{fig:results_se}(d) shows the final eccentricity. There are several observational estimates of the eccentricity, and it is known that the eccentricity of SENs can be approximately fitted by a Rayleigh distribution with $\sigma \sim 0.04-0.05$ \citep{2016PNAS..11311431X,2019AJ....157...61V,2019AJ....157..198M}. In our simulation results, the value is almost the same as the observational estimates. Although some small embryos with high eccentricity remain in the outer region $(a \gtrsim 2 {\rm \,au})$, SENs with $M > 1\,M_\oplus$ have small eccentricity $\lesssim 0.1$.

Figure~\ref{fig:results_se}(e) shows the relationship between eccentricity and multiplicity. Recent observations have shown that multiplicity and eccentricity are correlated, and that the higher the multiplicity, the lower the eccentricity \citep{2020AJ....160..276H,2024PSJ.....5..152L}. Our simulation results are in agreement with this observed trend. In our simulation results, the fraction of single systems is small. One of the reasons for this is that we stopped our simulations at $t=100 {\rm \,Myr}$, after which the fraction of single systems is expected to increase to some extent due to further dynamical instability.

We also comment on planetary composition. Although the composition of observed SENs is not strongly constrained due to degeneracy, it has been suggested that many of the SENs with $M \lesssim 10\,M_\oplus$ can be rocky compositions \citep[][]{2015ApJ...801...41R,2017ApJ...847...29O,2019PNAS..116.9723Z,2019ApJ...883...79D}. Although we do not directly follow the compositional evolution in this study, we expect that SENs have rocky composition, since they formed from rings that formed around $a=1\,{\rm au}$. We plan to investigate the actual compositional in a separate study using a planet formation model that takes compositional evolution into account. In the scenario of rapid migration in power-law disks, the formed SENs are expected to have a water-rich composition \citep[e.g.,][]{2020A&A...643L...1V,2021A&A...650A.152I}. Restrictions on the SEN formation model will be possible from further observational constraints on its composition.

\section{Discussion}\label{sec:discussion}

\begin{figure}[ht!]
\plotone{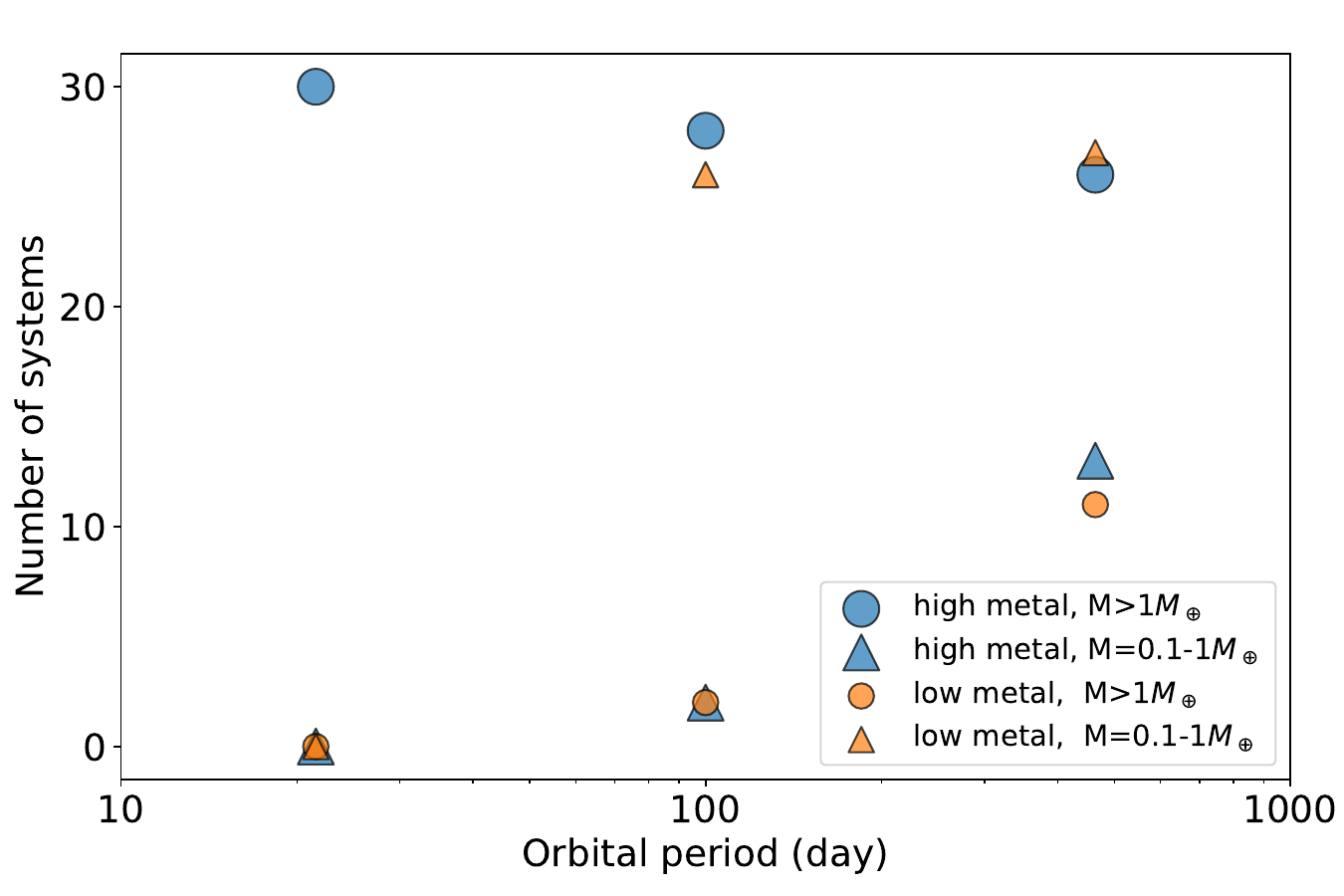}
\caption{Number of systems with planets in specific mass bins and orbital bins. Results of SEN formation with an initial total mass of $M_{\rm tot} = 20\,M_\oplus$ are indicated by blue symbols, while results of solar system formation with an initial total mass of $M_{\rm tot} = 2\,M_\oplus$ are indicated by orange symbols. The former is referred to as ``high metal'' and the latter as ``low metal.'' In each case we performed 30 systems with different $\overline{\alpha_{r\phi}}$ values. Planetary mass bins are divided into SENs with $M>1\,M_\oplus$ or sub-Earths with $M=0.1-1\,M_\oplus$. Orbital period bins are divided into three equal logarithmic bins between 10--1000 days.
\label{fig:occurrence}}
\end{figure}

\subsection{Occurrence of Earths and sub-Earths in the close-in region}
In our simulations, planets grow from rings of solids near $a=1\,{\rm au}$, and low-mass planets experience convergent/suppressed migration, while SENs experience slow inward migration. As a result, smaller planets than SENs are less likely to form in the close-in regions (Fig.~\ref{fig:results_se}(a)); in other words, this model predicts that hot Earths and hot sub-Earths are less common than hot SENs\footnote{On the other hand, if the inward migration of smaller planets is efficient, as in the case of the power-law disk, many Earths and sub-Earths can form in the close-in region (e.g., gray symbols in Fig.~\ref{fig:results_ss}(a)).}. Figure~\ref{fig:occurrence} shows the number of simulations (total 30 simulations each) that form specific planets in specific orbits. It can be seen that systems with sub-Earths in close-in orbits are rare in our simulations.

If observations show that the occurrence of these smaller terrestrial planets (Earths and sub-Earths) is lower than that of SENs in the close-in region, this could provide evidence in support of our formation model. Although we do not know for sure whether hot Earths and hot sub-Earths have lower occurrence than hot SENs based on current observations, there are observational data analysis studies that support this. \citet{2019AJ....158..109H} and \citet{2021AJ....161..201Q} used Kepler DR25 to estimate the occurrence rates of SENs and smaller planets. They found that in the close-in region, there is a peak in the occurrence of planets around the size of $R=1.4\,R_\oplus$ $(M \simeq 4\,M_\oplus)$, and that the occurrence of smaller planets decreases rapidly. We expect that future observations will reveal more details about the difference in occurrence between hot Earths/sub-Earths and hot SENs.

In Fig.~\ref{fig:results_se}(a) and Fig.~\ref{fig:occurrence}, the feature of no planets with $M\lesssim 1\,M_\oplus$ in the close-in region is clear, but in reality there should be more variation in the mass distribution. In our simulation, the initial total mass is set to $M_{\rm tot} = 20\,M_\oplus$ and the initial position of the ring is also fixed. In addition, there may be other disk profiles than the one we considered in Fig.~\ref{fig:disk}(a) that would realize the present scenario (slow inward migration for SENs and no migration for terrestrial planets). Simulations that vary these settings will give a little more variety in the mass distribution.

Note that \citet{2021AJ....161..201Q} pointed out the possibility that the occurrence increases for planets even smaller than $R=1\,R_\oplus$ (i.e., there is a gap in the occurrence around $R=1\,R_\oplus$). They explain this bimodality between SENs and sub-Earths by considering that sub-Earths are formed by a different mechanism than SENs. This idea is also consistent with our formation scenario in this paper.

\subsection{Planet-metallicity correlation}
Our results show that when the initial solid mass is large, SENs $(M \simeq 1-10\,M_\oplus)$ form in the close-in region, and SENs and Earths/sub-Earths also form in the region near $a=1\,{\rm au}$ (Fig.~\ref{fig:results_se}). On the other hand, when the initial solid mass is small, Earths/sub-Earths formed in the 1-au region do not migrate much, resulting in no planets forming in the close-in region (Fig.~\ref{fig:results_ss}). Assuming that the initial solid mass is correlated with the stellar metal abundance, there would be a correlation between planet occurrence and stellar metallicity. Note that the mass of the ring also depends on the disk viscosity, duration of the infall, etc \citep{2023NatAs...7..330B,2023A&A...677A.136M}. Therefore, the correlation between the ring mass and stellar metallicity is not simple. In addition, we have not done simulations with widely varying initial solid masses, and we have changed the gas disk model little; therefore, we cannot make a statistical argument. With this in mind, we discuss the planet-metallicity relation here.

As can be seen in Fig.~\ref{fig:occurrence}, the occurrence of SENs is more dependent on the initial total mass in the close-in region. In other words, the occurrence of close-in SENs is higher for high metallicity stars. This is a general result of our model. Other features that are less clear include the following. Figure~\ref{fig:occurrence} shows that the occurrence of SENs near $a=1\,{\rm au}$ is metallicity-dependent, but the metallicity dependence may be weaker than for the close-in SENs. In addition, there may be an inverse metallicity dependence for the occurrence of Earths and sub-Earths near $a=1\,{\rm au}$. This means that Earths and sub-Earths are more abundant in low-metallicity systems.

Previous observations have investigated the metallicity dependence of the occurrence of SENs in the close-in region \citep{2012Natur.486..375B,2015ApJ...799L..26S,2015AJ....149...14W,2018haex.bookE.153M,2018AJ....155...89P,2018PNAS..115..266D,2020AJ....159..247D}. They show that for SENs $(R=1-4\,R_\oplus)$ in the close-in region $(P=1-100\,{\rm day})$, their occurrence seems to correlate to some extent with the metallicity \citep{2015AJ....149...14W,2018AJ....155...89P}. Especially in the very close-in region of $P=1-10\,{\rm day}$, the correlation between the occurrence of SENs and metallicity seems to be large \citep{2018PNAS..115..266D,2018AJ....155...89P}. This is in agreement with our result.

The metallicity dependence of planet formation is certainly useful in constraining the planet formation model. However, again, our simulations do not make a strong claim because the initial ring mass can be changed by factors other than stellar metallicity and we do not perform simulations that consider a variety of situations. Further study of both theoretical models and observations will help to clarify the formation of low-mass planets.

\subsection{Various disk conditions}
The main purpose of this paper is to propose a way to solve the contradiction in solar system formation and SENs formation. Therefore, we show simulation results under limited conditions. Specifically, for the gas disk model, we consider the case where there is a density peak at $a \simeq 1\,{\rm au}$ at $t \simeq 1\,{\rm Myr}$. Although we perform simulations with slightly different value for $\overline{\alpha_{r\phi}}$ in Section~\ref{sec:characteristics}, some readers may wonder how our results might change with larger parameter changes. We should probably do the simulations over a wider range of parameters in another paper, but we briefly discuss this point.

First, for $\overline{\alpha_{r\phi}}$, we have considered the disk when increasing/decreasing by a factor of three from the reference value of $10^{-4}$ in Section~\ref{sec:characteristics}. As mentioned in Section~\ref{sec:model}, this value would be reasonable since low turbulence is favored. On the other hand, if the value of $\overline{\alpha_{r\phi}}$ is larger, the disk profile would change. As can be seen from Fig.~\ref{fig:disk}(a) and Fig.~1 of \citet{2016A&A...596A..74S}, the disk profile tends not to show clear density peaks. In this case, it is still the case that SENs migrate inward to form in the close-in region. However, Earths/sub-Earths can also experience some inward migration. This differs from our result (Fig.~\ref{fig:results_ss}) and makes it somewhat difficult to explain the radially concentrated orbits of the solar system's terrestrial planets. In this sense, we can say that peaked density disks are preferred for the solar system formation. In such a case, Earths/sub-Earths can migration into the close-in region and the picture in Fig.~\ref{fig:occurrence} would change quantitatively.

Next, for $\overline{\alpha_{\phi z}}$, which defines the strength of the wind-driven accretion, we use the relation of $\overline{\alpha_{\phi z}}=\min[10^{-5} (\Sigma_{\rm g}/\Sigma_{\rm g,ini})^{-0.66},1]$.
This means that the magnetic field does not dissipate as the disk dissipates, and the wind-driven accretion evolves to become larger with time. This causes the density peak to become more pronounced with time, as seen in Fig.~\ref{fig:disk}(a). The evolution of the gas surface density profile would change if the surface density dependence of $\overline{\alpha_{\phi z}}$ is changed. For example, if $\overline{\alpha_{\phi z}}$ does not change much with time, the surface density slope should be flatter than in Fig.~\ref{fig:disk}(a). An extreme example of this is shown as ``constant toqrque'' in Fig.~5 of \citet{2016A&A...596A..74S}, where the wind torque is constant with time.
In this case, as in the case of large $\overline{\alpha_{r\phi}}$ described above, the disk does not have a peaked density profile, and thus the convergent/suppressed migration of Earths/sub-Earths is unlikely to be realized. Note that if the gas surface density dependence of $\overline{\alpha_{\phi z}}$ is only slightly changed (the index is changed from -0.66 to -0.6), we confirm that the result does not change much.

In summary, if the disk density evolution used in this study is realized even if the disk parameters are changed, then our results are unchanged. On the other hand, if the disk profile is significantly different from the one we used, we would expect to see differences, especially in the Earths/sub-Earths migration. In this sense, if the occurrence of Earths and sub-Earths in the close-in region is further revealed by future observations, it will help to discuss the validity of our proposed formation model.

\section{Conclusions}\label{sec:conclusions}

While simulations of low-mass ($M \lesssim 10\,M_\oplus$) planet formation have been vigorously conducted over the past decade or so, we seem to face a contradiction. Simulations for the solar system's terrestrial planets require convergent/suppressed migration, while those for SENs require some degree of inward migration. However, simulations for the formation of the solar system's terrestrial planets do not produce SENs, and vice versa.
In this study, we investigate planet formation from a ring of solids in an evolving disk that develops a density peak at $r\simeq 1\,{\rm au}$ at $t\simeq 1\,{\rm Myr}$, and find that our results resolve the above contradiction. A summary of our findings is given below.
\begin{itemize}
    \item When low-mass planets of Earth or sub-Earth mass (corresponding to the solar system's terrestrial planets) form, they do not migrate inward due to the peaked disk profile. Note that the planets migrate within Mercury's orbit in power-law disks.
    \item The simulation results can reproduce some characteristics of the solar system's terrestrial planets. For example, the mass distribution of the inverted V-shape (Fig.~\ref{fig:results_ss}(a)) and metrics such as RMC and AMD (Fig.~\ref{fig:results_ss}(b)) are consistent with the present solar system.
    \item When SEN-mass planets grow out of rings, the planets as a whole experience a slow inward migration even in the disk with a peaked disk profile, because the outward migration region is limited. In the peaked disk model, the migration speed becomes slower and slower as the planets move inward, and the migration will eventually stop without the disk edge being set.
    \item Our simulation results also reproduce the characteristics of observed SENs. For example, the mass distribution is flat for $a=0.1-1\,{\rm au}$ (Fig.~\ref{fig:results_se}(a)), and the occurrence appears to be flat for $P=10-300\,{\rm day}$ (Fig.~\ref{fig:results_se}(b)). The orbital separation (Fig.~\ref{fig:results_se}(c)), eccentricity distribution (Fig.~\ref{fig:results_se}(d)), and eccentricity-multiplicity correlation (Fig.~\ref{fig:results_se}(e)) are also reasonably consistent with observations.
    \item We use the evolving disk model in which the gas surface density profile evolves with time and develops a density peak at $r \simeq 1 \,{\rm au}$ at about $t=1\,{\rm Myr}$. The simulation of terrestrial planet formation produces a radially concentrated planet distribution similar to the case of previous studies using a fixed density profile. This is because the planet mass is too small to undergo migration before $t=1 \,{\rm Myr}$ when there is no peak at $r \simeq 1 \,{\rm au}$. In the SEN formation simulation, planets moves across the density peak, but in this case the gas surface density is reduced in the inner region and the planets migrate slowly.
    \item Our simulation results provide a prediction for the distribution of close-in exoplanets (Section~\ref{sec:discussion}): Earths and sub-Earths are unlikely to form by our mechanism in the close-in region, as shown in the mass distribution in Fig.~\ref{fig:results_se}(a). Future observations may validate this model by comparing the occurrence of Earths/sub-Earths with that of SENs.
    \item Our results also allow us to speculate on the correlation between planet occurrence and stellar metallicity. Although not a firm conclusion, it is likely that the occurrence of SENs in the close-in region depends to some extent on stellar metallicity. This needs to be discussed in future studies where simulations with a wider range of parameter values are performed.
\end{itemize}

\begin{acknowledgments}
We thank the anonymous referee for useful comments and 
suggestions.
M.O. is supported by the National Natural Science Foundation of China (Nos. 12250610186, 12273023).
A.M. acknowledges funding from the ERC project N. 101019380 ``HolyEarth''.
M.K. is supported by the JSPS KAKENHI (grant nos. 23H01227, 24K00654, and 24K07099) and thanks Observatoire de la C\^{o}te d'Azur for the hospitality during his long-term stay in Nice.
Numerical computations were in part carried out on PC cluster at the Center for Computational Astrophysics, National Astronomical Observatory of Japan.
\end{acknowledgments}

\vspace{5mm}






\bibliography{mybib}{}

\begin{thebibliography}{}
\expandafter\ifx\csname natexlab\endcsname\relax\def\natexlab#1{#1}\fi
\providecommand{\url}[1]{\href{#1}{#1}}
\providecommand{\dodoi}[1]{doi:~\href{http://doi.org/#1}{\nolinkurl{#1}}}
\providecommand{\doeprint}[1]{\href{http://ascl.net/#1}{\nolinkurl{http://ascl.net/#1}}}
\providecommand{\doarXiv}[1]{\href{https://arxiv.org/abs/#1}{\nolinkurl{https://arxiv.org/abs/#1}}}

\bibitem[{{Alexander} \& {Armitage}(2007)}]{2007MNRAS.375..500A}
{Alexander}, R.~D., \& {Armitage}, P.~J. 2007, \mnras, 375, 500, \dodoi{10.1111/j.1365-2966.2006.11341.x}

\bibitem[{{Batygin} \& {Morbidelli}(2023)}]{2023NatAs...7..330B}
{Batygin}, K., \& {Morbidelli}, A. 2023, Nature Astronomy, 7, 330, \dodoi{10.1038/s41550-022-01850-5}

\bibitem[{{Bitsch} \& {Kley}(2010)}]{2010A&A...523A..30B}
{Bitsch}, B., \& {Kley}, W. 2010, \aap, 523, A30, \dodoi{10.1051/0004-6361/201014414}

\bibitem[{{Bitsch} {et~al.}(2015){Bitsch}, {Lambrechts}, \& {Johansen}}]{2015A&A...582A.112B}
{Bitsch}, B., {Lambrechts}, M., \& {Johansen}, A. 2015, \aap, 582, A112, \dodoi{10.1051/0004-6361/201526463}

\bibitem[{{Bro{\v{z}}} {et~al.}(2021){Bro{\v{z}}}, {Chrenko}, {Nesvorn{\'y}}, \& {Dauphas}}]{2021NatAs...5..898B}
{Bro{\v{z}}}, M., {Chrenko}, O., {Nesvorn{\'y}}, D., \& {Dauphas}, N. 2021, Nature Astronomy, 5, 898, \dodoi{10.1038/s41550-021-01383-3}

\bibitem[{{Buchhave} {et~al.}(2012){Buchhave}, {Latham}, {Johansen}, {Bizzarro}, {Torres}, {Rowe}, {Batalha}, {Borucki}, {Brugamyer}, {Caldwell}, {Bryson}, {Ciardi}, {Cochran}, {Endl}, {Esquerdo}, {Ford}, {Geary}, {Gilliland}, {Hansen}, {Isaacson}, {Laird}, {Lucas}, {Marcy}, {Morse}, {Robertson}, {Shporer}, {Stefanik}, {Still}, \& {Quinn}}]{2012Natur.486..375B}
{Buchhave}, L.~A., {Latham}, D.~W., {Johansen}, A., {et~al.} 2012, \nat, 486, 375, \dodoi{10.1038/nature11121}

\bibitem[{{Chambers}(2001)}]{2001Icar..152..205C}
{Chambers}, J.~E. 2001, \icarus, 152, 205, \dodoi{10.1006/icar.2001.6639}

\bibitem[{{Cresswell} \& {Nelson}(2008)}]{2008A&A...482..677C}
{Cresswell}, P., \& {Nelson}, R.~P. 2008, \aap, 482, 677, \dodoi{10.1051/0004-6361:20079178}

\bibitem[{{Dai} {et~al.}(2019){Dai}, {Masuda}, {Winn}, \& {Zeng}}]{2019ApJ...883...79D}
{Dai}, F., {Masuda}, K., {Winn}, J.~N., \& {Zeng}, L. 2019, \apj, 883, 79, \dodoi{10.3847/1538-4357/ab3a3b}

\bibitem[{{Dai} {et~al.}(2020){Dai}, {Winn}, {Schlaufman}, {Wang}, {Weiss}, {Petigura}, {Howard}, \& {Fang}}]{2020AJ....159..247D}
{Dai}, F., {Winn}, J.~N., {Schlaufman}, K., {et~al.} 2020, \aj, 159, 247, \dodoi{10.3847/1538-3881/ab88b8}

\bibitem[{{Dattilo} {et~al.}(2023){Dattilo}, {Batalha}, \& {Bryson}}]{2023AJ....166..122D}
{Dattilo}, A., {Batalha}, N.~M., \& {Bryson}, S. 2023, \aj, 166, 122, \dodoi{10.3847/1538-3881/acebc8}

\bibitem[{{Dong} {et~al.}(2018){Dong}, {Xie}, {Zhou}, {Zheng}, \& {Luo}}]{2018PNAS..115..266D}
{Dong}, S., {Xie}, J.-W., {Zhou}, J.-L., {Zheng}, Z., \& {Luo}, A. 2018, Proceedings of the National Academy of Science, 115, 266, \dodoi{10.1073/pnas.1711406115}

\bibitem[{{Dra{\.z}kowska} {et~al.}(2016){Dra{\.z}kowska}, {Alibert}, \& {Moore}}]{2016A&A...594A.105D}
{Dra{\.z}kowska}, J., {Alibert}, Y., \& {Moore}, B. 2016, \aap, 594, A105, \dodoi{10.1051/0004-6361/201628983}

\bibitem[{{Dullemond} {et~al.}(2018){Dullemond}, {Birnstiel}, {Huang}, {Kurtovic}, {Andrews}, {Guzm{\'a}n}, {P{\'e}rez}, {Isella}, {Zhu}, {Benisty}, {Wilner}, {Bai}, {Carpenter}, {Zhang}, \& {Ricci}}]{2018ApJ...869L..46D}
{Dullemond}, C.~P., {Birnstiel}, T., {Huang}, J., {et~al.} 2018, \apjl, 869, L46, \dodoi{10.3847/2041-8213/aaf742}

\bibitem[{{Flaherty} {et~al.}(2017){Flaherty}, {Hughes}, {Rose}, {Simon}, {Qi}, {Andrews}, {K{\'o}sp{\'a}l}, {Wilner}, {Chiang}, {Armitage}, \& {Bai}}]{2017ApJ...843..150F}
{Flaherty}, K.~M., {Hughes}, A.~M., {Rose}, S.~C., {et~al.} 2017, \apj, 843, 150, \dodoi{10.3847/1538-4357/aa79f9}

\bibitem[{{Hansen}(2009)}]{2009ApJ...703.1131H}
{Hansen}, B. M.~S. 2009, \apj, 703, 1131, \dodoi{10.1088/0004-637X/703/1/1131}

\bibitem[{{He} {et~al.}(2020){He}, {Ford}, {Ragozzine}, \& {Carrera}}]{2020AJ....160..276H}
{He}, M.~Y., {Ford}, E.~B., {Ragozzine}, D., \& {Carrera}, D. 2020, \aj, 160, 276, \dodoi{10.3847/1538-3881/abba18}

\bibitem[{{Hsu} {et~al.}(2019){Hsu}, {Ford}, {Ragozzine}, \& {Ashby}}]{2019AJ....158..109H}
{Hsu}, D.~C., {Ford}, E.~B., {Ragozzine}, D., \& {Ashby}, K. 2019, \aj, 158, 109, \dodoi{10.3847/1538-3881/ab31ab}

\bibitem[{{Ida} \& {Lin}(2008)}]{2008ApJ...673..487I}
{Ida}, S., \& {Lin}, D.~N.~C. 2008, \apj, 673, 487, \dodoi{10.1086/523754}

\bibitem[{{Izidoro} {et~al.}(2021){Izidoro}, {Bitsch}, {Raymond}, {Johansen}, {Morbidelli}, {Lambrechts}, \& {Jacobson}}]{2021A&A...650A.152I}
{Izidoro}, A., {Bitsch}, B., {Raymond}, S.~N., {et~al.} 2021, \aap, 650, A152, \dodoi{10.1051/0004-6361/201935336}

\bibitem[{{Izidoro} {et~al.}(2022){Izidoro}, {Dasgupta}, {Raymond}, {Deienno}, {Bitsch}, \& {Isella}}]{2022NatAs...6..357I}
{Izidoro}, A., {Dasgupta}, R., {Raymond}, S.~N., {et~al.} 2022, Nature Astronomy, 6, 357, \dodoi{10.1038/s41550-021-01557-z}

\bibitem[{{Izidoro} {et~al.}(2017){Izidoro}, {Ogihara}, {Raymond}, {Morbidelli}, {Pierens}, {Bitsch}, {Cossou}, \& {Hersant}}]{2017MNRAS.470.1750I}
{Izidoro}, A., {Ogihara}, M., {Raymond}, S.~N., {et~al.} 2017, \mnras, 470, 1750, \dodoi{10.1093/mnras/stx1232}

\bibitem[{{Jankovic} {et~al.}(2022){Jankovic}, {Mohanty}, {Owen}, \& {Tan}}]{2022MNRAS.509.5974J}
{Jankovic}, M.~R., {Mohanty}, S., {Owen}, J.~E., \& {Tan}, J.~C. 2022, \mnras, 509, 5974, \dodoi{10.1093/mnras/stab3370}

\bibitem[{{Kunitomo} {et~al.}(2020){Kunitomo}, {Suzuki}, \& {Inutsuka}}]{2020MNRAS.492.3849K}
{Kunitomo}, M., {Suzuki}, T.~K., \& {Inutsuka}, S.-i. 2020, \mnras, 492, 3849, \dodoi{10.1093/mnras/staa087}

\bibitem[{{Lambrechts} {et~al.}(2019){Lambrechts}, {Morbidelli}, {Jacobson}, {Johansen}, {Bitsch}, {Izidoro}, \& {Raymond}}]{2019A&A...627A..83L}
{Lambrechts}, M., {Morbidelli}, A., {Jacobson}, S.~A., {et~al.} 2019, \aap, 627, A83, \dodoi{10.1051/0004-6361/201834229}

\bibitem[{{Laskar}(1997)}]{1997A&A...317L..75L}
{Laskar}, J. 1997, \aap, 317, L75

\bibitem[{{Lissauer} {et~al.}(2023){Lissauer}, {Batalha}, \& {Borucki}}]{2023ASPC..534..839L}
{Lissauer}, J.~J., {Batalha}, N.~M., \& {Borucki}, W.~J. 2023, in Astronomical Society of the Pacific Conference Series, Vol. 534, Protostars and Planets VII, ed. S.~{Inutsuka}, Y.~{Aikawa}, T.~{Muto}, K.~{Tomida}, \& M.~{Tamura}, 839, \dodoi{10.48550/arXiv.2311.04981}

\bibitem[{{Lissauer} {et~al.}(2014){Lissauer}, {Dawson}, \& {Tremaine}}]{2014Natur.513..336L}
{Lissauer}, J.~J., {Dawson}, R.~I., \& {Tremaine}, S. 2014, \nat, 513, 336, \dodoi{10.1038/nature13781}

\bibitem[{{Lissauer} {et~al.}(2024){Lissauer}, {Rowe}, {Jontof-Hutter}, {Fabrycky}, {Ford}, {Ragozzine}, {Steffen}, \& {Nizam}}]{2024PSJ.....5..152L}
{Lissauer}, J.~J., {Rowe}, J.~F., {Jontof-Hutter}, D., {et~al.} 2024, \psj, 5, 152, \dodoi{10.3847/PSJ/ad0e6e}

\bibitem[{{Liu} {et~al.}(2019){Liu}, {Lambrechts}, {Johansen}, \& {Liu}}]{2019A&A...632A...7L}
{Liu}, B., {Lambrechts}, M., {Johansen}, A., \& {Liu}, F. 2019, \aap, 632, A7, \dodoi{10.1051/0004-6361/201936309}

\bibitem[{{Makino}(1991)}]{1991ApJ...369..200M}
{Makino}, J. 1991, \apj, 369, 200, \dodoi{10.1086/169751}

\bibitem[{{Makino} \& {Aarseth}(1992)}]{1992PASJ...44..141M}
{Makino}, J., \& {Aarseth}, S.~J. 1992, \pasj, 44, 141

\bibitem[{{Marschall} \& {Morbidelli}(2023)}]{2023A&A...677A.136M}
{Marschall}, R., \& {Morbidelli}, A. 2023, \aap, 677, A136, \dodoi{10.1051/0004-6361/202346616}

\bibitem[{{Masset} {et~al.}(2006){Masset}, {Morbidelli}, {Crida}, \& {Ferreira}}]{2006ApJ...642..478M}
{Masset}, F.~S., {Morbidelli}, A., {Crida}, A., \& {Ferreira}, J. 2006, \apj, 642, 478, \dodoi{10.1086/500967}

\bibitem[{{Matsumura} {et~al.}(2017){Matsumura}, {Brasser}, \& {Ida}}]{2017A&A...607A..67M}
{Matsumura}, S., {Brasser}, R., \& {Ida}, S. 2017, \aap, 607, A67, \dodoi{10.1051/0004-6361/201731155}

\bibitem[{{Millholland} {et~al.}(2022){Millholland}, {He}, \& {Zink}}]{2022AJ....164...72M}
{Millholland}, S.~C., {He}, M.~Y., \& {Zink}, J.~K. 2022, \aj, 164, 72, \dodoi{10.3847/1538-3881/ac7c67}

\bibitem[{{Mills} {et~al.}(2019){Mills}, {Howard}, {Petigura}, {Fulton}, {Isaacson}, \& {Weiss}}]{2019AJ....157..198M}
{Mills}, S.~M., {Howard}, A.~W., {Petigura}, E.~A., {et~al.} 2019, \aj, 157, 198, \dodoi{10.3847/1538-3881/ab1009}

\bibitem[{{Morbidelli} {et~al.}(2022){Morbidelli}, {Bailli{\'e}}, {Batygin}, {Charnoz}, {Guillot}, {Rubie}, \& {Kleine}}]{2022NatAs...6...72M}
{Morbidelli}, A., {Bailli{\'e}}, K., {Batygin}, K., {et~al.} 2022, Nature Astronomy, 6, 72, \dodoi{10.1038/s41550-021-01517-7}

\bibitem[{{Morishima} {et~al.}(2010){Morishima}, {Stadel}, \& {Moore}}]{2010Icar..207..517M}
{Morishima}, R., {Stadel}, J., \& {Moore}, B. 2010, \icarus, 207, 517, \dodoi{10.1016/j.icarus.2009.11.038}

\bibitem[{{Mulders}(2018)}]{2018haex.bookE.153M}
{Mulders}, G.~D. 2018, in Handbook of Exoplanets, ed. H.~J. {Deeg} \& J.~A. {Belmonte}, 153, \dodoi{10.1007/978-3-319-55333-7_153}

\bibitem[{{Ogihara} {et~al.}(2015{\natexlab{a}}){Ogihara}, {Kobayashi}, {Inutsuka}, \& {Suzuki}}]{2015A&A...579A..65O}
{Ogihara}, M., {Kobayashi}, H., {Inutsuka}, S.-i., \& {Suzuki}, T.~K. 2015{\natexlab{a}}, \aap, 579, A65, \dodoi{10.1051/0004-6361/201525636}

\bibitem[{{Ogihara} {et~al.}(2018{\natexlab{a}}){Ogihara}, {Kokubo}, {Suzuki}, \& {Morbidelli}}]{2018A&A...612L...5O}
{Ogihara}, M., {Kokubo}, E., {Suzuki}, T.~K., \& {Morbidelli}, A. 2018{\natexlab{a}}, \aap, 612, L5, \dodoi{10.1051/0004-6361/201832654}

\bibitem[{{Ogihara} {et~al.}(2018{\natexlab{b}}){Ogihara}, {Kokubo}, {Suzuki}, \& {Morbidelli}}]{2018A&A...615A..63O}
---. 2018{\natexlab{b}}, \aap, 615, A63, \dodoi{10.1051/0004-6361/201832720}

\bibitem[{{Ogihara} {et~al.}(2020){Ogihara}, {Kunitomo}, \& {Hori}}]{2020ApJ...899...91O}
{Ogihara}, M., {Kunitomo}, M., \& {Hori}, Y. 2020, \apj, 899, 91, \dodoi{10.3847/1538-4357/aba75e}

\bibitem[{{Ogihara} {et~al.}(2015{\natexlab{b}}){Ogihara}, {Morbidelli}, \& {Guillot}}]{2015A&A...578A..36O}
{Ogihara}, M., {Morbidelli}, A., \& {Guillot}, T. 2015{\natexlab{b}}, \aap, 578, A36, \dodoi{10.1051/0004-6361/201525884}

\bibitem[{{Owen} {et~al.}(2012){Owen}, {Clarke}, \& {Ercolano}}]{2012MNRAS.422.1880O}
{Owen}, J.~E., {Clarke}, C.~J., \& {Ercolano}, B. 2012, \mnras, 422, 1880, \dodoi{10.1111/j.1365-2966.2011.20337.x}

\bibitem[{{Owen} \& {Wu}(2017)}]{2017ApJ...847...29O}
{Owen}, J.~E., \& {Wu}, Y. 2017, \apj, 847, 29, \dodoi{10.3847/1538-4357/aa890a}

\bibitem[{{Paardekooper} {et~al.}(2010){Paardekooper}, {Baruteau}, {Crida}, \& {Kley}}]{2010MNRAS.401.1950P}
{Paardekooper}, S.~J., {Baruteau}, C., {Crida}, A., \& {Kley}, W. 2010, \mnras, 401, 1950, \dodoi{10.1111/j.1365-2966.2009.15782.x}

\bibitem[{{Paardekooper} {et~al.}(2011){Paardekooper}, {Baruteau}, \& {Kley}}]{2011MNRAS.410..293P}
{Paardekooper}, S.~J., {Baruteau}, C., \& {Kley}, W. 2011, \mnras, 410, 293, \dodoi{10.1111/j.1365-2966.2010.17442.x}

\bibitem[{{Paardekooper} \& {Papaloizou}(2009{\natexlab{a}})}]{2009MNRAS.394.2283P}
{Paardekooper}, S.~J., \& {Papaloizou}, J.~C.~B. 2009{\natexlab{a}}, \mnras, 394, 2283, \dodoi{10.1111/j.1365-2966.2009.14511.x}

\bibitem[{{Paardekooper} \& {Papaloizou}(2009{\natexlab{b}})}]{2009MNRAS.394.2297P}
---. 2009{\natexlab{b}}, \mnras, 394, 2297, \dodoi{10.1111/j.1365-2966.2009.14512.x}

\bibitem[{{Pan} {et~al.}(2024){Pan}, {Liu}, {Johansen}, {Ogihara}, {Wang}, {Ji}, {Wang}, {Feng}, \& {Ribas}}]{2024A&A...682A..89P}
{Pan}, M., {Liu}, B., {Johansen}, A., {et~al.} 2024, \aap, 682, A89, \dodoi{10.1051/0004-6361/202347664}

\bibitem[{{Petigura} {et~al.}(2018){Petigura}, {Marcy}, {Winn}, {Weiss}, {Fulton}, {Howard}, {Sinukoff}, {Isaacson}, {Morton}, \& {Johnson}}]{2018AJ....155...89P}
{Petigura}, E.~A., {Marcy}, G.~W., {Winn}, J.~N., {et~al.} 2018, \aj, 155, 89, \dodoi{10.3847/1538-3881/aaa54c}

\bibitem[{{Qian} \& {Wu}(2021)}]{2021AJ....161..201Q}
{Qian}, Y., \& {Wu}, Y. 2021, \aj, 161, 201, \dodoi{10.3847/1538-3881/abe632}

\bibitem[{{Raymond} {et~al.}(2009){Raymond}, {O'Brien}, {Morbidelli}, \& {Kaib}}]{2009Icar..203..644R}
{Raymond}, S.~N., {O'Brien}, D.~P., {Morbidelli}, A., \& {Kaib}, N.~A. 2009, \icarus, 203, 644, \dodoi{10.1016/j.icarus.2009.05.016}

\bibitem[{{Rogers}(2015)}]{2015ApJ...801...41R}
{Rogers}, L.~A. 2015, \apj, 801, 41, \dodoi{10.1088/0004-637X/801/1/41}

\bibitem[{{Schlaufman}(2015)}]{2015ApJ...799L..26S}
{Schlaufman}, K.~C. 2015, \apjl, 799, L26, \dodoi{10.1088/2041-8205/799/2/L26}

\bibitem[{{Suzuki} \& {Inutsuka}(2014)}]{2014ApJ...784..121S}
{Suzuki}, T.~K., \& {Inutsuka}, S.-i. 2014, \apj, 784, 121, \dodoi{10.1088/0004-637X/784/2/121}

\bibitem[{{Suzuki} {et~al.}(2010){Suzuki}, {Muto}, \& {Inutsuka}}]{2010ApJ...718.1289S}
{Suzuki}, T.~K., {Muto}, T., \& {Inutsuka}, S.-i. 2010, \apj, 718, 1289, \dodoi{10.1088/0004-637X/718/2/1289}

\bibitem[{{Suzuki} {et~al.}(2016){Suzuki}, {Ogihara}, {Morbidelli}, {Crida}, \& {Guillot}}]{2016A&A...596A..74S}
{Suzuki}, T.~K., {Ogihara}, M., {Morbidelli}, A., {Crida}, A., \& {Guillot}, T. 2016, \aap, 596, A74, \dodoi{10.1051/0004-6361/201628955}

\bibitem[{{Tabone} {et~al.}(2022){Tabone}, {Rosotti}, {Cridland}, {Armitage}, \& {Lodato}}]{2022MNRAS.512.2290T}
{Tabone}, B., {Rosotti}, G.~P., {Cridland}, A.~J., {Armitage}, P.~J., \& {Lodato}, G. 2022, \mnras, 512, 2290, \dodoi{10.1093/mnras/stab3442}

\bibitem[{{Taki} {et~al.}(2021){Taki}, {Kuwabara}, {Kobayashi}, \& {Suzuki}}]{2021ApJ...909...75T}
{Taki}, T., {Kuwabara}, K., {Kobayashi}, H., \& {Suzuki}, T.~K. 2021, \apj, 909, 75, \dodoi{10.3847/1538-4357/abd79f}

\bibitem[{{Thiemens} {et~al.}(2019){Thiemens}, {Sprung}, {Fonseca}, {Leitzke}, \& {M{\"u}nker}}]{2019NatGe..12..696T}
{Thiemens}, M.~M., {Sprung}, P., {Fonseca}, R. O.~C., {Leitzke}, F.~P., \& {M{\"u}nker}, C. 2019, Nature Geoscience, 12, 696, \dodoi{10.1038/s41561-019-0398-3}

\bibitem[{{Touboul} {et~al.}(2007){Touboul}, {Kleine}, {Bourdon}, {Palme}, \& {Wieler}}]{2007Natur.450.1206T}
{Touboul}, M., {Kleine}, T., {Bourdon}, B., {Palme}, H., \& {Wieler}, R. 2007, \nat, 450, 1206, \dodoi{10.1038/nature06428}

\bibitem[{{Ueda} {et~al.}(2019){Ueda}, {Flock}, \& {Okuzumi}}]{2019ApJ...871...10U}
{Ueda}, T., {Flock}, M., \& {Okuzumi}, S. 2019, \apj, 871, 10, \dodoi{10.3847/1538-4357/aaf3a1}

\bibitem[{{Ueda} {et~al.}(2021){Ueda}, {Ogihara}, {Kokubo}, \& {Okuzumi}}]{2021ApJ...921L...5U}
{Ueda}, T., {Ogihara}, M., {Kokubo}, E., \& {Okuzumi}, S. 2021, \apjl, 921, L5, \dodoi{10.3847/2041-8213/ac2f3b}

\bibitem[{{Van Eylen} {et~al.}(2019){Van Eylen}, {Albrecht}, {Huang}, {MacDonald}, {Dawson}, {Cai}, {Foreman-Mackey}, {Lundkvist}, {Silva Aguirre}, {Snellen}, \& {Winn}}]{2019AJ....157...61V}
{Van Eylen}, V., {Albrecht}, S., {Huang}, X., {et~al.} 2019, \aj, 157, 61, \dodoi{10.3847/1538-3881/aaf22f}

\bibitem[{{Venturini} {et~al.}(2020){Venturini}, {Guilera}, {Haldemann}, {Ronco}, \& {Mordasini}}]{2020A&A...643L...1V}
{Venturini}, J., {Guilera}, O.~M., {Haldemann}, J., {Ronco}, M.~P., \& {Mordasini}, C. 2020, \aap, 643, L1, \dodoi{10.1051/0004-6361/202039141}

\bibitem[{{Wang} \& {Fischer}(2015)}]{2015AJ....149...14W}
{Wang}, J., \& {Fischer}, D.~A. 2015, \aj, 149, 14, \dodoi{10.1088/0004-6256/149/1/14}

\bibitem[{{Weiss} \& {Marcy}(2014)}]{2014ApJ...783L...6W}
{Weiss}, L.~M., \& {Marcy}, G.~W. 2014, \apjl, 783, L6, \dodoi{10.1088/2041-8205/783/1/L6}

\bibitem[{{Weiss} {et~al.}(2023){Weiss}, {Millholland}, {Petigura}, {Adams}, {Batygin}, {Block}, \& {Mordasini}}]{2023ASPC..534..863W}
{Weiss}, L.~M., {Millholland}, S.~C., {Petigura}, E.~A., {et~al.} 2023, in Astronomical Society of the Pacific Conference Series, Vol. 534, Protostars and Planets VII, ed. S.~{Inutsuka}, Y.~{Aikawa}, T.~{Muto}, K.~{Tomida}, \& M.~{Tamura}, 863

\bibitem[{{Wimarsson} {et~al.}(2020){Wimarsson}, {Liu}, \& {Ogihara}}]{2020MNRAS.496.3314W}
{Wimarsson}, J., {Liu}, B., \& {Ogihara}, M. 2020, \mnras, 496, 3314, \dodoi{10.1093/mnras/staa1708}

\bibitem[{{Wolfgang} {et~al.}(2016){Wolfgang}, {Rogers}, \& {Ford}}]{2016ApJ...825...19W}
{Wolfgang}, A., {Rogers}, L.~A., \& {Ford}, E.~B. 2016, \apj, 825, 19, \dodoi{10.3847/0004-637X/825/1/19}

\bibitem[{{Woo} {et~al.}(2023){Woo}, {Morbidelli}, {Grimm}, {Stadel}, \& {Brasser}}]{2023Icar..39615497W}
{Woo}, J.~M.~Y., {Morbidelli}, A., {Grimm}, S.~L., {Stadel}, J., \& {Brasser}, R. 2023, \icarus, 396, 115497, \dodoi{10.1016/j.icarus.2023.115497}

\bibitem[{{Xie} {et~al.}(2016){Xie}, {Dong}, {Zhu}, {Huber}, {Zheng}, {De Cat}, {Fu}, {Liu}, {Luo}, {Wu}, {Zhang}, {Zhang}, {Zhou}, {Cao}, {Hou}, {Wang}, \& {Zhang}}]{2016PNAS..11311431X}
{Xie}, J.-W., {Dong}, S., {Zhu}, Z., {et~al.} 2016, Proceedings of the National Academy of Science, 113, 11431, \dodoi{10.1073/pnas.1604692113}

\bibitem[{{Zeng} {et~al.}(2019){Zeng}, {Jacobsen}, {Sasselov}, {Petaev}, {Vanderburg}, {Lopez-Morales}, {Perez-Mercader}, {Mattsson}, {Li}, {Heising}, {Bonomo}, {Damasso}, {Berger}, {Cao}, {Levi}, \& {Wordsworth}}]{2019PNAS..116.9723Z}
{Zeng}, L., {Jacobsen}, S.~B., {Sasselov}, D.~D., {et~al.} 2019, Proceedings of the National Academy of Science, 116, 9723, \dodoi{10.1073/pnas.1812905116}

\end{thebibliography}
\bibliographystyle{aasjournal}



\end{document}